\documentclass[aps,prl,twocolumn,groupedaddress]{revtex4-1}

\usepackage{graphicx}
\usepackage{xcolor}
\usepackage{float}
\usepackage[caption = false]{subfig}

\begin{document}

\title{Forgetting Memories and their Attractiveness}

\author{Enzo Marinari}
\email[]{enzo.marinari@uniroma1.it}
\affiliation{Sapienza Universit\'a di Roma, INFN Sezione di Roma 1
  and Nanotech-CNR, UOS di Roma, P.le A. Moro 2, 00185 Roma, Italy}

\date{\today}

\begin{abstract}
We study numerically the memory which forgets, introduced in 1986 by
Parisi by bounding the synaptic strength, with a mechanism which avoid
confusion, allows to remember the pattern learned more recently and
has a physiologically very well defined meaning.  We analyze a number
of features of the learning at finite number of neurons and finite
number of patterns.  We discuss how the system behaves in the large
but finite $N$ limit. We analyze the basin of attraction of the
patterns that have been learned, and we show that it is exponentially
small in the age of the pattern. This is a clearly non physiological
feature of the model.
\end{abstract}


\maketitle

The model proposed by Hopfield \cite{hopfield_1982} in 1982,
implementing in a statistical mechanical setting ideas from Hebb
\cite{hebb_book_1949} and Eccles \cite{eccles_book_1953} has surely
been a dramatic breakthrough in the study of neural systems. In the
last years a huge amount of work has helped to try and bring these
ideas closer to realistic systems (see for example one of
\cite{murest_book_1995,itmiro_edited_1997,rollstreves_book_1998,
  rolls_book_1999,engvan_book_2001,engste_book_2004,rollsdeco_book_2010,
  kaswje_book_2013,rolls_book_2014,rolls_book_2016} and of references
therein).

Hopfield model can learn a number of patterns that grows linearly in
the number of neurons $N$, This is a welcome feature but it goes with
a serious drawback. When one tries to have the system learning more
than this maximal fraction of patterns the system enters a state of
complete confusion, and forgets everything; no patterns can be learned
in this situation.

Parisi solved this problem in 1986 with an elegant proposal
\cite{parisi_1986}, noticing that bounding the strength of the
synapses is enough to allow the memory to forget. Now less patterns
are remembered than in the original Hopfield model (but always with a
capacity proportional to $N$), but if new patterns are shown to the
system it forgets the old patterns and learns the new ones. It is
clear that physiologically the synaptic strength cannot grow
indefinitely, that gives in this sense to Parisi approach a strong
physiological common sense. Parisi model is in principle based on
synapses that must be known with high precision: this is a strong
physiological weakness, since it is experimentally clear that no more
than order of ten levels of synaptic intensity can be relevant (see
for example \cite{benfus_2016} and references therein). Models where a
lower synaptic accuracy is needed can be important, but we will not
deal with them here.  The model proposed by Parisi has been solved
analytical, in the limit of $N\to\infty$ in \cite{vakeku_1988}. Also
other variations (the so called palimpsestic approaches) have been
introduced in \cite{natoch_1986} (where a threshold is also used) and
in \cite{menato_1986}. These are basic models that need to be largely
elaborated and modified in order to become good potential descriptions
of realistic systems. A lot of efforts have been done in this
direction (see for example \cite{fussen_2006,fusabb_2007,benfus_2016}
and references therein).

In this note we will try to clarify some important features of the
basic Parisi approach, since this is an important basis of many recent
developments. We will analyze numerically detailed features of the
learning at finite number of neurons and finite number of patterns,
also as a function, for example, of the number of patterns that have
been shown to the system. We will show how the system behaves in the
large but finite $N$ limit. We will analyze the basin of attraction of
the recognizable patterns, and detect what seems to be a clearly
non physiological feature of the model. 

\textit{Model and Parameters} --- In our model we have $N$ neurons
$\sigma_i$, $i=1,... N$ that can take the values $\pm 1$. We will
denote a configuration of the $N$ neurons by $\{\sigma\}$. The model
is of the mean field type since in principle each neuron $i$ is
connected to all neurons $j$ by a synaptic strength $J_{i,j}$. The
synaptic connections are formed by learning from patterns. A learning
rule makes the synaptic connections dependent on $M$ patterns
(collections of $N$ values $\tau_i=\pm 1$ that we denote by
$\{\tau\}$). We call $R=M/N$ the ratio of presented patterns over
number of neurons.

The dynamics of the neurons is dictated by the synapses dependent
energy function:
\begin{equation}
\label{eENE}
  E\left[\sigma\right]=-\frac12 \sum_{i\ne j} \sigma_i\, J_{i,j}\,\sigma_j\;,
\end{equation}
where positive synaptic couplings enhance the affinity among a given
couple of neurons, while negative couplings suppress it. When a
pattern $\{\tau\}$ is presented to the system the synapses evolve. We
write this evolution as:
\begin{equation}
\label{eRULE}
  J_{i,j}^{\mbox{new}} = f\left(  J_{i,j}^{\mbox{old}} +
  \frac{\tau_i\tau_j}{\sqrt{N}}  \right)\;,
\end{equation}
There are $M$ patterns that in the model are presented, one by one,
only one time each, to the system. After the learning the system could
be able to recognize some of the patterns that have been
presented. Finding a pattern is defined here as a situation where
after energy minimization the system ends close enough one of the $M$
patterns that have been presented. By recognition we mean starting
from one of the pattern presented and used to form the synapses, and
after energy minimization ending close enough to this same pattern (by
close enough we mean that in the final pattern no more than $\epsilon
N$ neurons differ from the ones of the starting pattern: in all this
note we take $\epsilon=0.02$). We define the recovery rate $\rho$ as
the number of recognized patterns over the number of neurons $N$.
It is very interesting and telling, as we will see, to compute the
basin of attraction of the different stored patterns, by starting, for
example, from a random point in phase space (or in a sphere around one
of the presented patterns).

The original generalized Hebb is given by the simple linear relation
$f(x) = x$. Hebb rule is in some sense very successful, since it
allows to store and recognize a number of patterns proportional to
$N$. If the number of patterns one wants to store $M$, is small enough
($M<\alpha_c N\sim 0.14N$), this can turn out to be what one needs,
but when $M$ is larger than the critical threshold $\alpha$ a disaster
occurs, since the memory gets completely confused, and everything is
forgotten. A very natural and simple mechanism for solving this
problem has been proposed in \cite{parisi_1986,natoch_1986}, and, in
the limit of $M\to\infty$ and $N\to\infty$, the Parisi model
\cite{parisi_1986} has been solved analytically in \cite{vakeku_1988}.
Here synapses cannot grow more than a constant saturation threshold
$A$. So $f(x)=x$ if $\;-A< x< A$ (adopting in this regime the original
generalized Hebb rule), but $f(x)=-A$ for $x\le -A$ and $f(x)=A$ for
$x\ge A$. Bounding the values of the synapses is physiologically very
reasonable (even if more features have to be calibrated in order to
get a realistic model, see for example \cite{benfus_2016} and
references therein): the structure of this forgetting model is in this
sense both very simple and very reasonable. In his $1986$ numerical
simulations Parisi estimated that the model does not learn for
$A>A_c\sim0.7$ (a very precise estimate since the exact solution gives
$A_c\sim .692$). On the contrary for small values of $A$ the model can
have a finite storage capability, close to $0.04$ (i.e. smaller than
the one of the Hopfield model).

\begin{figure}
\subfloat{(a)}{\includegraphics[width=0.9\columnwidth]{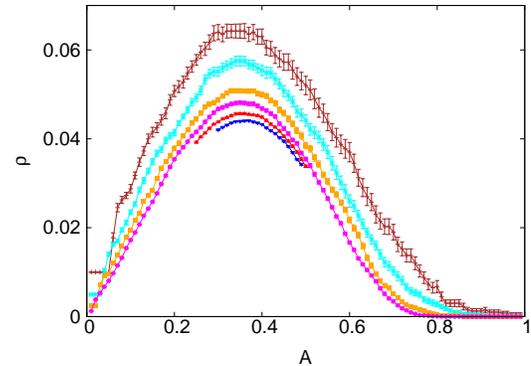}}
\subfloat{(b)}{\includegraphics[width=0.9\columnwidth]{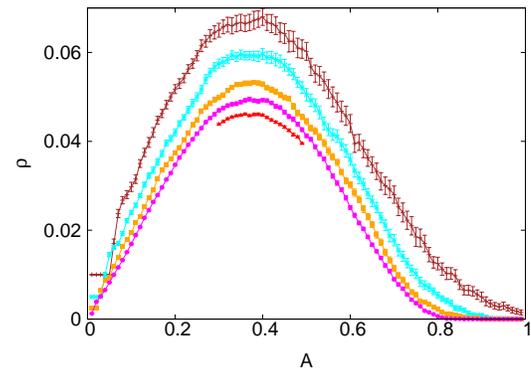}}
\caption{Recognition rate as a function of $A$
  for $R=1$ (a) and $R=2$ (b). From top to bottom $N=100$, 200,
  400, 800, 1600 and $N=3200$ for $R=1$ only.\label{fREC}}
\end{figure}

\textit{The Recovery Rate as a Function of the Threshold $A$} --- As a
first step we have measured the recognition rate $\rho$ as a function
of $A$. Given the couplings that have been learned from patterns we
start, one by one, from all patterns that have been shown to the
system, minimize the energy function (\ref{eENE}) by changing the
neurons $\{\sigma\}$ in a steepest descent, and check if the arrival
point is close enough (with a precision $\epsilon$, as we have
explained before) to the starting point. We use values of $N$ going
from 100 to 3200, values of $M$ such to have ratios $R=M/N$ going from
1 to 8, We average over $40$ different realizations of the set of
patterns but for the case $N=3200$, $M=3200$ where we have $20$
samples and the case $N=1600$, $M=3200$ where we have $30$ samples.
Statistical errors are computed as fluctuations over the independent
samples.  In the Supplemental Material (SM), available online, we show
all the available data and plot them in different ways. We show in
Fig. \ref{fREC} the recovery rate as a function of the threshold $A$
for $R=1$ and for $R=2$ (asymptotically for large $N$ and $M$ the
system remembers in both case, and independently from $R$, the last
$N$ patterns shown: see later). The two plots show clear finite $N$
effects, and a comparison of the two also shows that the value of $M$
plays a role. In the upper plot one shows one pattern per neuron, and
the asymptotic result is closer than in the case where we show to the
system two pattern for each neuron: finite size effects increase with
$R$. The large $A$ region where $\rho=0$ is very clear, and it is
moving to lower values of $A$ for increasing values of $N$. For large
$N$ $\rho$ departs linearly from $A=0$, increases linearly in a
finite range of $A$, has a maximum and eventually reaches zero.  The
value of $A$ where the recovery rate is maximal, $A_{\mbox{max}}$ is
well defined already at these values of $N$. 

Fig. (\ref{fREC}) makes clear that already for a finite (and
reasonably large) number of patterns the behavior of the system is
close to the asymptotic behavior (our systems with $N=3200$ neurons
can learn more than $120$ patterns), and that the details of the
learning also depend from the number of patterns presented to the
system (for finite values of $N$ and $M$).

\begin{figure}
\subfloat{(a)}{\includegraphics[width=0.9\columnwidth]{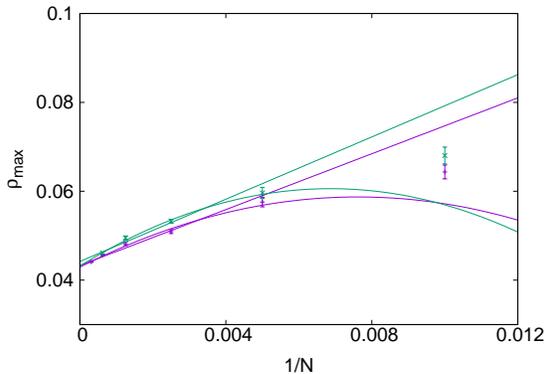}}
\subfloat{(b)}{\includegraphics[width=0.9\columnwidth]{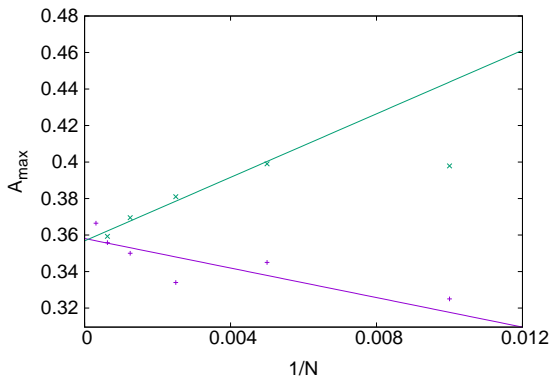}}
\caption{(a) Maximum recognition rate $\rho_{\mbox{max}}$ as a
  function of $1/N$ for $R=1$ (plus symbols on the lower curves) and
  for $R=2$ (multiplication signs on the upper curves). The straight
  lines are for the linear fits, the curved lines for the quadratic
  fits (see text).  (b) The value of the threshold $A_{\mbox{max}}$
  such that $\rho$ is maximum versus $1/N$ (symbols like in panel
  (a)), and best linear fits to the data.
  \label{fMAX}}
\end{figure}

\textit{The Optimal Value of the Threshold $A$} --- As a first step we
want to understand how the maximal recognition rate is reached in the
$N\to\infty$ limit at fixed values of $R$. In order to do that we have
first interpolated the maxima of the curves of Figs. (\ref{fREC}). We
have used a quadratic interpolation around the measured maximum
value. We plot in Fig. (\ref{fMAX}) the maximum recognition rate
$\rho_{\mbox{max}}$ and the value of the threshold $A_{\mbox{max}}$
such that $\rho$ is maximum, versus $1/N$. We fit the data for
$\rho_{\mbox{max}}$ in two different ways.

Both for $R=1$ and $R=2$, separately, we use a linear fit and a
quadratic fit, in both cases discarding the point with $N=100$. As it
is clear from panel (a) of the figure from all these fits we get
compatible results for the value of $\rho_{\mbox{max}}$ in the
$N\to\infty$ limit, where, for example, the discrepancy among the
values measured at finite $N$ for $R=1$ and $R=2$ shrinks (when using
the quadratic fit it becomes of the order of one part per thousand in
the $N\to\infty$ limit). The estimated asymptotic value is
$0.0431(5)$, and it is slightly larger (of less than 5\%) than the
theoretical value quoted by \cite{vakeku_1988}, maybe because of a
pronounced curvature in the $N\to\infty$ limit.

Our estimates of $A_{\mbox{max}}$ are less precise, so we only use
linear fits, and we show them in panel (b) of Fig. (\ref{fMAX}); also
here we discard the $N=100$ data. Remarkably the finite $N$
corrections have opposite signs in the two cases $R=1$ and $R=2$, and
the finite $N$ optimal values of the threshold are very different, but
in the $N\to\infty$ limit for fixed $R$ the two sets of data converge
to very similar values. Averaging the two values we estimate
$A_{\mbox{max}}^{N=\infty}=.357(5)$, that has to be compared to the
theoretical values $.354$ \cite{vakeku_1988}. The fact that when
different (large) numbers of patterns have been shown to the system
the optimal learning threshold can vary is an important feature we are
learning here.

\begin{figure}
\includegraphics[width=0.9\columnwidth]{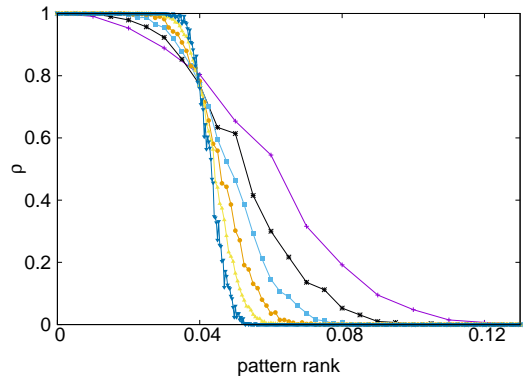}
\caption{$\rho$ as a function of the pattern rank, i.e. of the
  position in which a pattern has been shown to the system divided by
  $M$ (on the left are the most recent patterns, on the right the
  oldest ones: the pattern rank goes from 0 to 1). $A=0.40$,
  very close to the threshold value where the recognition rate is
  maximum, $R=1$ and (from top to bottom in the right part of the
  figure) $N=100$, 200, 400, 800, 1600 and 3200. $S=1000$ samples for
  all cases but 2000 samples for $N=1600$ and 250 samples for
  $N=3200$.
  \label{fTHETA}}
\end{figure}

\textit{The Recovery Rate as a Function of the Pattern Rank} --- As we
have discussed the remarkable feature of our model is thta it can
forget. When new patterns are learned old patterns are forgotten: the
storage capacity is dedicated to the patterns learned more
recently. From an intuitive point of view it is not clear if in the
$N\to\infty$ limit all this capacity strictly moves to the last
patterns shown to the system. Here we analyze this issue from a
numerical, finite $N$ point of view. In Fig. (\ref{fTHETA}) we show
$\rho$ as a function of the pattern rank $r$ for systems with $R=1$
and different number of neurons.  Here $A=4$, where the recognition
rate is high.

\begin{figure}
\includegraphics[width=0.9\columnwidth]{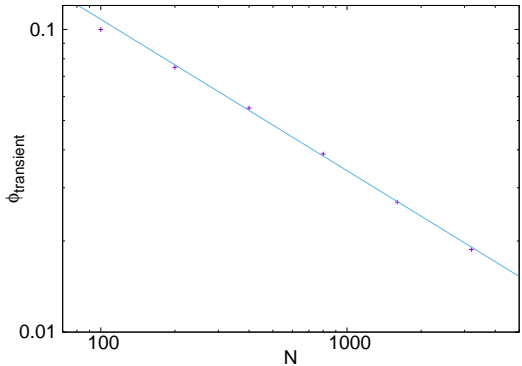}
\caption{The fraction of recognized patterns in the transient region
  $\phi_{\mbox{transient}}$,
  as defined in the text, as a function of $N$. Here $R=1$ and
  $A=0.40$.
  The continuous line is for the function $1.08\, x^{-0.5}$.
  \label{fFRACTION}}
\end{figure}

For increasing $N$ the jump from the recognized region to the
forgotten region becomes sharper: already on our larger systems, here
with $N=M=3200$, there is a very clear jump. In order to make
quantitative the visual evidence that the recognition rate $\rho$ is
becoming a step function we analyze the region where the recognition
rate is smaller than one and larger then zero, and we check if it is
shrinking to zero for $N\to\infty$. We look at the ordered patterns
(starting with the most recently shown that have smaller rank) and we
have the transient region starting when we observe the first pattern
with a recognition rate smaller than $1-\eta$ (we set
$\eta=0.01$). The transient region ends when we meet the first pattern
with a recognition rate smaller than $\eta$, and we compute the
fraction of patterns $\phi_{\mbox{transient}}$ in the transient
region, that we show in Fig. (\ref{fFRACTION}).

$\phi_{\mbox{transient}}$ goes to zero when $N\to\infty$. We plot in
the figure the function $1.08\, x^{-0.5}$, that describes perfectly the
data. We have shown a clear, unambiguous evidence that the 
recognition rate as a function of the pattern rank becomes a
$\theta$--function in the $N\to\infty$ limit.

\begin{figure}
\subfloat{(a)}{\includegraphics[width=0.9\columnwidth]{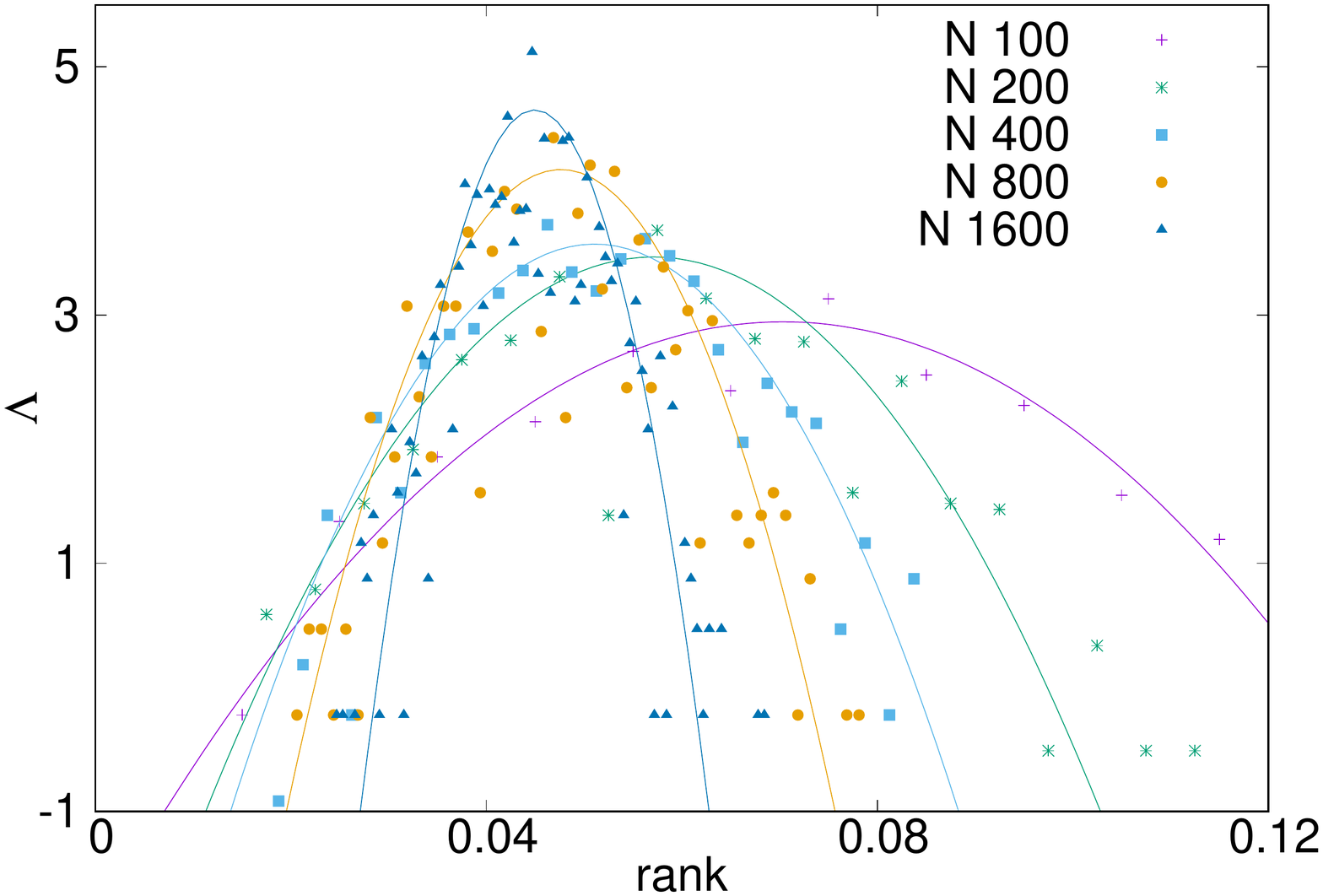}}
\subfloat{(b)}{\includegraphics[width=0.9\columnwidth]{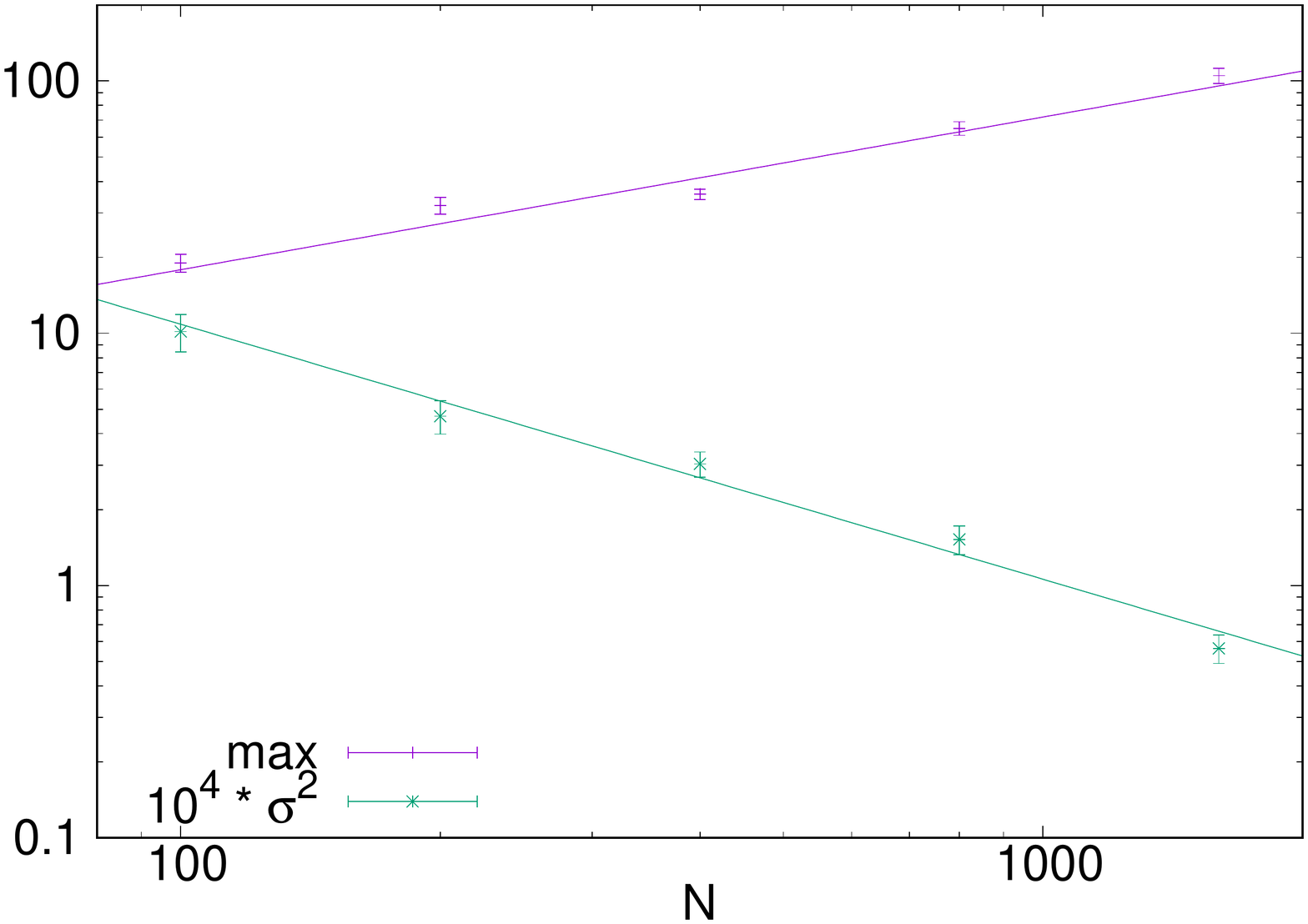}}
\caption{(a) $\Lambda$ versus rank for different values of $N$, $R=1$
  and $A=0.4$. (b) The maximum of the Gaussian functions used to fit
  $\Lambda(r)$ at fixed $N$ and their widths as a function of $N$. The
  straight lines are for the best fits to a power law. 
  \label{fDELTA}}
\end{figure}

It is useful to look at the logarithm of minus the derivative of the
recognition rate with respect to the rank
$$
\Lambda\equiv - \log\left(\frac{d\rho}{d\mbox{(rank)}}\right)\;.
$$
We show it in panel (a) of Fig. (\ref{fDELTA}) for $R=1$, $A=0.4$ and
different values of $N$ (our statistics for systems with $N=M=3200$ is
not good enough to be useful in this analysis).

The continuous line in  Fig. (\ref{fDELTA}) are for a best fit of
$\Lambda$ to the logarithm of a Gaussian function, i.e. we use a three
parameters fits in n$c$, $\tilde{r}$ and $ \sigma_\Lambda$:
$$
\Lambda(r) \sim c - \frac{\left(r-\tilde{r}\right)^2}{\sigma^2_\Lambda}\; .
$$
The best fits are very good, and reconstruct the data with remarkable
accuracy. The Gaussian functions become narrow when $N$ increases, and
the heights of their maxims increase.

To make this statement quantitative we look at the dependence over $N$
of the maximum of the Gaussian functions and at their width, and we
plot them, in log-log scale, in panel (b) of Fig.  (\ref{fDELTA}). The
straight lines are here for our best power law fits to the values of
the peak and of $\sigma^2_\Lambda$, that describe very accurately the
data. We find that the maximum of $\Lambda$ grows as $N$ to the power
$0.6\pm 0.1$, while $\sigma^2_\Lambda$ decreases when $N$ increases
with a power $-1.0\pm 0.1$. The evidence for the recognition rate
being a $\theta$ function in the $N\to\infty$ limit is clear.

\textit{The Basin of Attraction of the Recognized Patterns} --- We are
considering systems with $N$ neurons, where after showing $M$ random,
uncorrelated patterns we have been able to ``learn'' some of these
patterns. A very relevant question is about the basin of attraction of
these patterns that we are able to recognize (as we have seen, the
last that have been shown to the system). We will discuss the answer
to a precise, simple question. If we start from one random pattern,
extracted with uniform probability among all the available neuron
configurations, how probable it is that we ``recognize'' one of the
patterns that we have stored in our memory? Some of the starting
patterns will lead to no recognition, and some will lead instead to
``recognize'' one of the patterns that have been stored in the
system. In a ``reasonable'' physical systems all stored (random and
uncorrelated) patterns should have a similar probability of being
retrieved one starting from a random pattern: this probability could
mildly decrease with the age of pattern (i.e. with the time that has
passed after its presentation to the system), but should not collapse
for older patterns. As we will see, and we find this to be a strong
inherent weakness of the system we are studying,this is not true here.

\begin{figure}
\includegraphics[width=0.9\columnwidth]{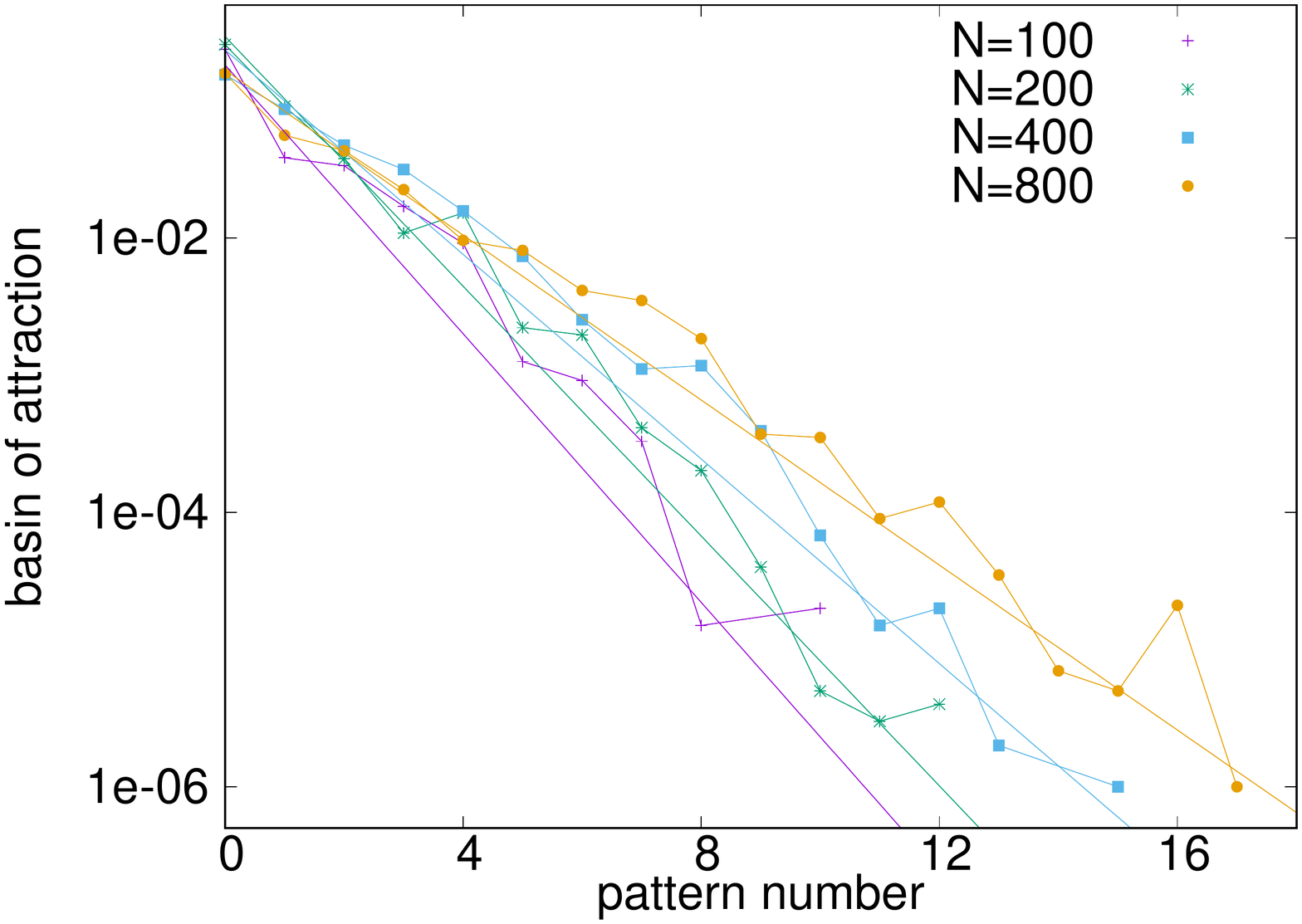}
\caption{The full basin of attraction of recognized patterns (i.e. the
  percentage of starts from a random point in phase space that lead to
  a given stored pattern, as a function of the pattern number ($0$ is
  for the last pattern presented to the system).
  \label{fBASIN}}
\end{figure}

We show in Fig. (\ref{fBASIN}) the basin of attraction of recognized
patterns (defined as we have just discussed: we start from a random
point in phase space, we have the energy decrease by steepest descent,
and check if we hit a pattern at distance closer or equal to
$\epsilon$, using, as before $\epsilon=0.02$) as a function of the
pattern number.  For all $N$ values that we show in Fig.
(\ref{fBASIN}) we start from $10^5$ random patterns for each of the
$10$ different sets of couplings we analyze (samples), and we average
over the ten samples.  For all $N$ values where we have statistically
significant results (these are the ones we show in
Fig. (\ref{fBASIN})) there is a clear exponential decrease of the
basin of the attraction of the pattern as a function of the age of the
pattern: we are able to observe a decrease of the order of $10^5$.
Old patterns are only very rarely recognized: they are ``stored'',
since starting from the exact pattern one falls very close to the
pattern itself, but recognizing them when starting not too close is
very difficult.  The exponential decay of the basins of attraction
with the time from presentation is probably related to exponential
degradation for increasing noise observed
\cite{benfus_2016,fusabb_2007}.  We are using a here a very simple
learning schedule, where patterns are only shown ones: a more complex
schedule with repetitions could probably alleviate this problems, but
the capacity of the system would decrease, and, crucially,
physiologically this does not look like an appropriate mechanism.

The probability that starting from a random pattern the system does
not reach any of the stored pattern does not change dramatically with
$N$: it is 0.67 for $N=100$, 0.58 for $N=200$, 0.65 for $N=400$ and
0.69 for $N=800$. This confirms the idea that observing the basin of
convergence to a recognized pattern when starting from a random
pattern is a sensible measure of how a recognized pattern is
accessible when the systems tries to recognize.

\textit{Conclusions} --- We have analyzed a simple and basic model
(proposed by Parisi \cite{parisi_1986} and solved analytically in the
thermodynamical limit by van Hemmen, Keller and K\"uhn
\cite{vakeku_1988}) of a memory that can forget. We have analyzed and
unveiled in detail, numerically, its behavior for finite number of
neurons $N$ and finite number of patterns shown to the system, $M$,
and its dependence over $R\equiv M/N$. We have been able to answer
numerically about issues like the concentration of the recognizable
pattern on the most recently learned patterns in the limit of $N$,
$M\to\infty$ (where the recognizable pattern are always and only that
last that have been learned). We have found that the size of the basin
of attraction of the recognized patterns decreases exponentially with
the time distance from the moment when the pattern has been
presented. This feature is not reasonable from a physiological point
of view, and should be cured when trying to describe realistically
neural systems: it is possible that modifications of palimpsestic
schemes \cite{natoch_1986,menato_1986} could be useful in this
context.

\begin{acknowledgments}
  
I thank for the nice hospitality the Santa Barbara Kavli Institute for
Theoretical Physics (KITP) and its 2018 \textit{Memory Formation in
  Matter} programme, where the most part of the present work was
developed.  I thank Arvidn Murugan for interesting discussions, and I
am deeply grateful to Stefano Fusi and to Marcus Benna for very useful
comments on a first version of the manuscript.  This research was
supported in part by the National Science Foundation under Grant
No. NSF PHY17-48958, and by the European Research Council (ERC) under
the European Union’s Horizon 2020 Research and Innovation Program
(grant agreement n. 694925).

\end{acknowledgments}

\bibliography{em_memory}

\end{document}